\documentclass[12pt]{article}

\newcommand{\be}{\begin{equation}}
\newcommand{\ee}{\end{equation}}

\begin{document}

\begin{center}
{\large Letters in Mathematical Physics, 
Vol.73. No.1. (2005) pp.49-58.}
\end{center}

\begin{center}
{\Large \bf Fractional Generalization of Gradient Systems}
\end{center}

\vskip 5mm
\begin{center}
{\large \bf Vasily E. Tarasov } \\
{\it Skobeltsyn Institute of Nuclear Physics, \\ 
Moscow State University, Moscow 119992, Russia \\
E-mail: tarasov@theory.sinp.msu.ru } \\
\end{center}

\begin{abstract}
We consider a fractional generalization of gradient systems. 
We use differential forms 
and exterior derivatives of fractional orders. 
Examples of fractional gradient systems are considered. 
We describe the stationary states of these systems. 
\end{abstract}


\vskip 5mm

\section{Introduction}

Derivatives and integrals of fractional order \cite{SKM,OS,Podlubny} 
have found many applications in recent studies in physics.
The interest in fractional analysis has been growing continually 
during the last few years. 
Fractional analysis has numerous applications: 
kinetic theories \cite{Zaslavsky1,Zaslavsky2,Physica};  
statistical mechanics \cite{chaos,PRE05,JPCS}; 
dynamics in complex media \cite{Nig,PLA05,PLA05-2,AP05,Chaos05}; 
and many others.

The theory of derivatives of non-integer order goes back 
to Leibniz, Liouville, Riemann, Grunwald, and Letnikov. 
In the last few decades many authors have pointed out that 
fractional-order models are more appropriate than integer-order 
models for various real materials. 
Fractional derivatives provide an excellent instrument 
for the description of memory and hereditary properties 
of various materials and processes. 
This is the main advantage of fractional derivatives 
in comparison with classical integer-order models in 
which such effects are in fact neglected. 
The advantages of fractional derivatives become apparent 
in modelling mechanical and electrical properties of real materials, 
as well as in the description of rheological properties of rocks, 
and in many other fields. 

In this Letter, we use
a fractional generalization of exterior calculus that was suggested
in Refs. \cite{FDF,FDF2}. Fractional generalizations of differential forms
and exterior derivatives were defined in \cite{FDF}.
It allows us to consider the fractional generalization
of gradient dynamical systems \cite{Gilmor,DNF}.
The suggested class of fractional gradient systems
is a wider class than the class of usual gradient dynamical systems. 
The gradient systems can be considered as
special case of fractional gradient systems. 

In Section 2, a brief review of 
gradient systems and exterior calculus 
is considered to fix notation and provide a convenient reference.
In Section 3, a definition of fractional generalization 
of gradient systems is suggested.
In Section 4, we consider a fractional gradient system that
cannot be considered as a gradient system. 
In Section 5, we prove that a dynamical system that 
is defined by the well-known Lorenz equations \cite{Lor1,Lor2} 
can be considered as a fractional gradient system.
Finally, a short conclusion is given in Section 6.

\section{Gradient Systems}

In this section, a brief review of 
gradient systems and exterior calculus \cite{DNF}
is considered to fix notation and provide a convenient reference.

Gradient systems arise in dynamical systems theory \cite{HS,DNF,Gilmor}. 
They are described by the equation 
$d{\bf x}/dt=- grad \ V(x)$, where ${\bf x} \in R^n$.
In Cartesian coordinates, the gradient is given by
$grad \ V={\bf e}_i {\partial V}/{\partial x_i}$,
where ${\bf x}={\bf e}_i x_i$. Here and later, we mean the sum on 
the repeated indices $i$ and $j$ from 1 to n. \\

{\bf Definition 1.}
{\it A dynamical system that is described by the equations
\be  \label{cs} \frac{dx_i}{dt}=F_i(x)  \quad (i=1,...,n) \ee
is called a gradient system in $R^n$ if the differential 1-form 
\be \label{omega} \omega=F_i(x)dx_i \ee
is an exact form $\omega=-dV$, where
$V=V(x)$ is a continuously differentiable function (0-form). } \\

Here, $d$ is the exterior derivative \cite{DNF}.
Let $V=V(x)$ be a real, continuously differentiable function on $R^n$.
The exterior derivative of the function $V$ is the one form
$dV=dx_i {\partial V}/{\partial x_i}$ 
written in a coordinate chart $(x_1,...,x_n)$.

In mathematics \cite{DNF}, 
the concepts of closed form and exact form are defined 
for differential forms, by the equation $d\omega =0$
for a given form $\omega$ to be a closed form,
and $\omega=dh$ for an exact form.
It is known, that to be exact is a sufficient condition to be closed. 
In abstract terms the question of whether this is also a necessary condition 
is a way of detecting topological information, by differential conditions.

Let us consider the 1-forms (\ref{omega}). 
The formula for the exterior derivative $d$ of (\ref{omega}) is
\[ d \omega=\frac{1}{2} \left(\frac{\partial F_i}{\partial x_j} - 
\frac{\partial F_j}{\partial x_i} \right) dx_j \wedge dx_i , \]
where $\wedge$ is the wedge product. 
Therefore the condition for $\omega$ to be closed is
\be \label{FxFx} \frac{\partial F_i}{\partial x_j} - 
\frac{\partial F_j}{\partial x_i}=0. \ee
In this case, if $V(x)$ is a potential function then
$dV = dx_i{\partial V}/{\partial x_i}$.
The implication from 'exact' to 'closed' is then a consequence 
of the symmetry of the second derivatives commute, 
\be \label{Vxy} \frac{\partial^2 V}{\partial x_i \partial x_j}=
\frac{\partial^2 V}{\partial x_j \partial x_i} . \ee
If the function $V=V(x)$ is smooth function, then the second derivative
commute, and Eq. (\ref{Vxy}) holds. \\

{\bf Proposition 1.}
{\it If a smooth vector field ${\bf F}={\bf e}_i F_i(x)$ of system (\ref{cs})
satisfies the relations (\ref{FxFx})
on a contractible open subset X of $R^n$,
then the dynamical system (\ref{cs}) is
the gradient system such that }
\be \label{ps} \frac{dx_i}{dt}=-\frac{\partial V(x)}{\partial x_i} . \ee

If the exact differential 1-form $\omega$ is equal to zero
($dV=0$), then we get the equation $V(x)-C=0$
that defines the stationary states of gradient dynamical 
system (\ref{ps}). Here, $C$ is a constant.

\section{Fractional Generalization of Gradient Systems}

A fractional generalization of exterior calculus was suggested
in \cite{FDF,FDF2}. A fractional exterior derivative 
and fractional differential forms were defined \cite{FDF}.
It allows us to consider the fractional generalization
of gradient systems.
 
If the partial derivatives in the definition of the exterior derivative 
$d=dx_i \partial / \partial x_i$ are allowed to assume fractional order, 
a fractional exterior derivative can be defined \cite{FDF} by the equation
$d^{\alpha}=(dx_i)^{\alpha} {\bf D}^{\alpha}_{x_i}$. 
Here, we use the fractional derivative ${\bf D}^{\alpha}_{x}$
in the Riemann-Liouville form \cite{SKM} that is defined by the equation
\be \label{df} {\bf D}^{\alpha}_{x}f(x)=\frac{1}{\Gamma (m-\alpha)}
\frac{\partial^m}{\partial x^m}  
\int^x_{0} \frac{f(y) dy}{(x-y)^{\alpha-m+1}} , \ee
where $m$ is the first whole number greater than  or equal to $\alpha$.
The initial point of the fractional derivative \cite{SKM} is set to zero.
The derivative of powers $k$ of $x$ is
\be \label{xk} {\bf D}^{\alpha}_x x^k=
\frac{\Gamma(k+1)}{\Gamma(k+1-\alpha)} x^{k-\alpha} ,\ee
where $k \ge 1$, and $\alpha \ge 0$. 
Note that the derivative of a constant $C$ need not be zero. 

Let us consider a dynamical system  that is defined
by the equation ${d{\bf x}}/{dt}={\bf F}$, 
on subset X of $R^n$. 
In Cartesian coordinates, we can use Eq. (\ref{cs}),
where $i=1,..,n$, ${\bf x}=x_i{\bf e}_i$, and ${\bf F}=F_i{\bf e}_i$.
The fractional analog of Definition 1 has the form. \\

{\bf Definition 2.} 
{\it A dynamical system (\ref{cs})
is called a fractional gradient system if the 
fractional differential 1-form 
\be \label{omega-a} \omega_{\alpha}=F_i(x)(dx_i)^{\alpha} \ee 
is an exact fractional form $\omega_{\alpha}=-d^{\alpha} V$, 
where $V=V(x)$ is a continuously differentiable function, and
$d^{\alpha}=(dx_i)^{\alpha} {\bf D}^{\alpha}_{x_i}$ 
is a fractional exterior derivative. } \\

Using the definition of the fractional exterior derivative, 
Eq. (\ref{omega-a}) can be represented as
\[ \omega_{\alpha}=-d^{\alpha} V=-(dx_i)^{\alpha} {\bf D}^{\alpha}_{x_i}V. \]
Therefore, we have $F_i(x)=-{\bf D}^{\alpha}_{x_i} V$. 

Note that Eq. (\ref{omega-a}) is a fractional generalization
of differential form (\ref{omega}). 
Obviously that fractional 1-form $\omega_{\alpha}$
can be closed when the differential 1-form $\omega=\omega_1$ is not closed. \\

{\bf Proposition 2.}
{\it If a smooth vector field ${\bf F}={\bf e}_i F_i(x)$ 
on a contractible open subset X of $R^n$ 
satisfies the relations
\be \label{FHC} {\bf D}^{\alpha}_{x_j} F_i- {\bf D}^{\alpha}_{x_i} F_j=0 , \ee
then the dynamical system (\ref{cs})
is a fractional gradient system such that 
\be \label{fps} \frac{dx_i}{dt}=-{\bf D}^{\alpha}_{x_i} V(x) , \ee
where $V(x)$ is a continuous differentiable function and 
${\bf D}^{\alpha}_{x_i}V=-F_i$ }. \\

{\bf Proof}. This proposition is a corollary of the 
fractional generalization of Poincare lemma \cite{FDF2}.
The Poincare lemma is shown \cite{FDF,FDF2}
to be true for exterior fractional derivative. \\

Note that the Riemann-Liouville fractional derivative of a 
constant need not be zero (\ref{xk}), and we have 
\[ {\bf D}^{\alpha}_{x_i} C=\frac{x^{-\alpha}_i}{\Gamma(1-\alpha)} C . \]
Therefore  we see that constants $C$ in the equation
$V(x)=C$ cannot define a stationary state for Eq. (\ref{fps}). 
In order to define stationary states of fractional gradient systems, 
we consider the solutions of system of the equations 
${\bf D}^{\alpha}_{x_i} V(x)=0$. \\

{\bf Proposition 3.} 
{\it The stationary states of gradient system (\ref{fps})
are defined by the equation 
\be \label{ssgs} V(x)-\left|\prod^n_{i=1} x_i \right|^{\alpha-m} 
\sum^{m-1}_{k_1=0} ... \sum^{m-1}_{k_n=0} C_{k_1...k_n} 
\Bigl( \prod^n_{i=1}(x_i)^{k_i} \Bigr)=0 . \ee
The $C_{k_1...k_n}$ are constants and $m$ is the first whole number
greater than  or equal to $\alpha$.}\\

{\bf Proof.} 
In order to define the stationary states of a fractional gradient system, 
we consider the solution of the equation
\be \label{DH} {\bf D}^{\alpha}_{x_i} V(x)=0 . \ee
This equation can be solved by using Eq. (\ref{df}).
Let $m$ be the first whole number greater than or equal to $\alpha$,
then we have the solution \cite{OS,SKM} of Eq. (\ref{DH}) in the form
\be \label{sDF} V(x)=|x_i|^{\alpha} \sum^{m-1}_{k=0} 
a_k(x_1,...,x_{i-1},x_{i+1},...,x_n) (x_i)^{k} , \ee
where the $a_k$ are functions of the other coordinates.
Using Eq. (\ref{sDF}) for $i=1,...,n$,  
we get the solution of the system of equation (\ref{DH})
in the form (\ref{ssgs}). \\

If we consider $n=2$ such that $x=x_1$ and $y=x_2$, we have
the equations of motion for fractional gradient system
\be \label{xH2} \frac{dx}{dt} =-{\bf D}^{\alpha}_{x} V(x,y), 
\quad \frac{dy}{dt} =-{\bf D}^{\alpha}_{y} V(x,y) .\ee
The stationary states for Eqs. (\ref{xH2})
are defined by the equation 
\[ V(x,y)-|xy|^{\alpha-1} \sum^{m-1}_{k=0} \sum^{m-1}_{l=0} C_{kl} 
x^{k} y^{l}=0 . \]
The $C_{kl}$ are constants and $m$ is the first whole number
greater than  or equal to $\alpha$.

The Riemann-Liouville fractional derivative has some notable 
disadvantages in physical applications such as 
the hyper-singular improper integral, where the order 
of singularity is higher than the dimension, and 
nonzero of the fractional derivative of constants, 
which would entail that dissipation does not vanish for a 
system in equilibrium.  
The desire to formulate initial value problems for physical
systems leads to the use of Caputo fractional derivatives \cite{Podlubny}
rather than Riemann-Liouville fractional derivative.
The Caputo fractional derivative 
\cite{Podlubny,Caputo,Caputo2,Mainardi} is defined by 
\be \label{Caputo} 
{\bf D}^{\alpha}_{* \ x}f(x)=\frac{1}{\Gamma (m-\alpha)}
\int^x_{0} \frac{f^{(m)}(y) dy}{(x-y)^{\alpha-m+1}} , \ee
where $f^{(m)}(y)=d^m f(y)/dy^m$, and
$m$ is the first whole number greater than or equal to $\alpha$.
This definition is of course more restrictive that (\ref{df}),
in that requires the absolute integrability of the derivative
of order $m$. 
The Caputo fractional derivative first computes an ordinary
derivative followed by a fractional integral to achieve the
desire order of fractional derivative.
The Riemann-Liouville fractional derivative 
is computed in the reverse order. 
Integration by part of (\ref{Caputo}) will lead to 
\be \label{C-RL}
{\bf D}^{\alpha}_{* \ x}f(x)={\bf D}^{\alpha}_{x}f(x)-
\sum^{m-1}_{k=0}
\frac{x^{k-\alpha}}{\Gamma(k-\alpha+1)} f^{(k)}(0+) .
\ee 
It is observed that the second term in Eq. (\ref{C-RL}) regularizes 
the Caputo fractional derivative to avoid the potentially divergence 
from singular integration at $x=0+$. In addition, the 
Caputo fractional differentiation of a constant results in zero.
 
If the Caputo fractional derivative is used instead of the 
Riemann-Liouville  fractional derivative then the stationary states 
of fractional gradient systems are the same as those for
the usual gradient systems ($V(x)-C=0$).  
The Caputo formulation of fractional calculus can be more 
applicable to gradient systems than the Riemann-Liouville formulation.

\section{Examples of Fractional Gradient System}

In this section, we consider fractional gradient systems that
cannot be considered as a gradient system. 
We prove that the class of fractional gradient systems
is a wider class than  the usual class of gradient dynamical systems. 
The gradient systems can be considered as
special case of fractional gradient systems. \\

{\bf Example 1.}
Let us consider the dynamical system that is defined by the equations
\be \label{eqex1} \frac{dx}{dt}=F_x, \quad \frac{dy}{dt}=F_y ,\ee
where the right hand sides of equations (\ref{eqex1}) have the form
\be \label{ex1} 
F_x=a c x^{1-k}+ b x^{-k} , \quad F_y=(ax+b) y^{-k} , \ee
where $a\not=0$. 
This system cannot be considered as a gradient dynamical system.
Using 
\[ \frac{\partial F_x}{\partial y}-\frac{\partial F_y}{\partial x}=
ay^{-k} \not=0 , \]
we get that $\omega=F_xdx+F_ydy$ is not closed form
\[ d\omega=-a y^{-k} dx \wedge dy .\]
Note that the relation (\ref{FHC}) in the form
\[ {\bf D}^{\alpha}_y F_x-{\bf D}^{\alpha}_x F_y=0 , \]
is satisfied for the system (\ref{ex1}), if $\alpha=k$ and 
the constant $c$ is defined by \ 
$c={\Gamma(1-\alpha)}/{\Gamma(2-\alpha)}$. 
Therefore, this system
can be considered as a fractional gradient system with 
$\alpha=k$ and a linear potential function
\[ V(x,y)=\Gamma(1-\alpha) (ax+b) .\]

{\bf Example 2.}
Let us consider the dynamical system that is defined 
by Eq. (\ref{eqex1}) with  
\[ F_x=an(n-1)x^{n-2}+ck(k-1)x^{k-2}y^l , \]
\[ F_y=bm(m-1)y^{m-2}+cl(l-1)x^ky^{l-2} , \]
where $k\not=1$ and $l\not=1$. It is easy to derive that 
\[ \frac{\partial F_x}{\partial y}-\frac{\partial F_y}{\partial x}=
ckl \ x^{k-2} y^{l-2} ((k-1)y-(l-1)x) \not=0 ,  \]
and the differential form $\omega=F_xdx+F_ydy$ is not closed $d \omega\not=0$.
Therefore this system is not a gradient dynamical system. 
Using the conditions (\ref{FHC}) in the form
\[ {\bf D}^2_y F_x-{\bf D}^2_x F_y=
\frac{\partial^2 F_x}{\partial y^2}-
\frac{\partial^2 F_y}{\partial x^2}=0 ,  \]
we get $d^{\alpha} \omega=0$ for $\alpha=2$. 
As the result, we have that this system can be considered as a fractional 
gradient system with $\alpha=2$ and the potential function
\[ V(x,y)=ax^n+by^m+cx^ky^l . \]

In the general case, the fractional gradient system cannot
be considered as gradient system. 
The gradient systems can be considered 
as special case of fractional gradient systems such that $\alpha=1$.

\section{Lorenz System as a Fractional Gradient System}

In this section, we prove that dynamical systems that 
are defined by the well-known Lorenz equations \cite{Lor1,Lor2} 
are fractional gradient systems. \\

The well-known Lorenz equations \cite{Lor1,Lor2} 
are defined by 
\[ \frac{dx}{dt}=F_x ,\quad 
\frac{dy}{dt}=F_y , \quad 
\frac{dz}{dt}=F_z , \]
where the right hand sides $F_x$, $F_y$ and $F_z$ have the forms
\[ F_x=\sigma (y-x) ,\quad F_y=(r-z)x-y , \quad  F_z=xy-bz . \]
The parameters $\sigma$, $r$ and $b$ can be equal to the following 
values
\[ \sigma=10, \quad b=8/3, \quad r=470/19 \simeq 24.74 \ . \]
The dynamical system which is defined by the Lorenz equations
cannot be considered as gradient dynamical system. 
It is easy to see that
\[ \frac{\partial F_x}{\partial y}-\frac{\partial F_y}{\partial x}
=z+\sigma-r ,\]
\[ \frac{\partial F_x}{\partial z}-\frac{\partial F_z}{\partial x}
=-y , \quad \quad \quad \quad \]
\[ \frac{\partial F_y}{\partial z}-\frac{\partial F_z}{\partial y}
=-2x . \quad \quad\quad \]
Therefore $\omega=F_xdx+F_ydy+F_zdz$ is not a closed 1-form and we have 
\[ d \omega=-(z+\sigma-r)dx \wedge dy+y dx \wedge dz+
2x dy \wedge dz . \]

For the Lorenz equations, the conditions (\ref{FHC}) are satisfied:
\[ {\bf D}^2_y F_x-{\bf D}^2_x F_y=0,
\quad {\bf D}^2_z F_x-{\bf D}^2_x F_z=0,
\quad {\bf D}^2_z F_y-{\bf D}^2_y F_z=0 . \]
As the result, we get that the Lorenz system can be considered 
as a fractional gradient dynamical system with potential function
\be \label{LP} V(x,y,z)=\frac{1}{6}\sigma x^3-\frac{1}{2}\sigma yx^2
+\frac{1}{2}(z-r)xy^2+ \frac{1}{6} y^3 
-\frac{1}{2}xyz^2+\frac{b}{6}z^3 . \ee
The potential (\ref{LP}) uniquely defines the Lorenz system.
Using Eq. (\ref{ssgs}), we can get that the stationary states of 
the Lorenz system are defined by $\alpha=m=2$, and  the equation
\be \label{LSS} V(x,y,z)+
C_{00}+C_x x+C_y y+C_z z+C_{xy}xy+C_{xz}xz+C_{yz}yz=0, \ee
where $C_{00}$, $C_x$, $C_y$, $C_z$ $C_{xy}$, $C_{xz}$, and $C_{yz}$
are the constants. 
The plot of these stationary states of Lorenz system
can be derived by using computer.

Note that 
the Rossler system \cite{Ros}, that is defined by the equations
\[ \frac{dx}{dt}=-(y+z), \quad 
\frac{dx}{dt}=x+0.2y, \quad \frac{dz}{dt}=0.2+(x-c)z ,\] 
can be considered as a fractional gradient system with 
potential function
\[ V(x,y,z)=\frac{1}{2}(y+z)x^2-\frac{1}{2}xy^2-
\frac{1}{30} y^3 -\frac{1}{10}z^2-\frac{1}{6}(x-c)z^3 . \]
This potential uniquely defines the Rossler system. 
The stationary states of the Rossler system 
are defined by Eq. (\ref{LSS}).

\section{Conclusion}

Using the fractional derivatives and fractional differential
forms, we consider the fractional generalization
of gradient systems.
In the general case, the fractional gradient system cannot
be considered as gradient systems. 
The class of fractional gradient systems
is a wider class than the usual class of gradient dynamical systems. 
The gradient systems can be considered as
special case of fractional gradient systems. 
Therefore, it is possible to generalize the application 
of catastrophe and bifurcation theory from
gradient to a wider class of fractional gradient dynamical systems. 
Note that the order of fractional derivative can be considered
as an additional parameter that can leads to bifurcation.
For example, fractional gradient system with 
the Ginzburg-Landau potential 
\[ V(x)=\frac{1}{4}x^4+\frac{a}{2}x^2+bx  \]
has the stationary states that are defined by the equation
$x^{1-\alpha}(x^3+a'x+b')=0$,
where bifurcation is defined by new values of parameters
\[ a'=a\frac{\Gamma(5-\alpha)}{6\Gamma(3-\alpha)},
\quad b'=b\frac{\Gamma(5-\alpha)}{6\Gamma(2-\alpha)} . \]

Note also that the some of fractional gradient systems
can be non-local in coordinates, due to the integral in the
definition of fractional derivatives.
These non-local properties will be considered in the next paper.
The fractional generalization of differential forms \cite{FDF,FDF2} 
leads us to the following open questions:
Is there a fractional analog of the homology and cohomology theories?
Is there a connection with the non-local character 
of the fractional derivative and the topological properties 
of the fractional differential forms?  
These interesting open questions require the additional research.

\end{document}